\documentclass[12pt]{article}
\usepackage{graphicx, amsmath, amssymb, cite, setspace, color, relsize, bm, bbold, array, hyperref, dsfont, bbold, xcolor, cancel,soul}

\usepackage[top=1 in, bottom=.82 in, left=.9 in, right=.9 in]{geometry}

\newcommand\T{\rule{0pt}{2.5ex}}       
\newcommand\B{\rule[-1.2ex]{0pt}{0pt}} 

\newcommand{\captionfonts}{\footnotesize} 
\makeatletter
\long\def\@makecaption#1#2{%
  \vskip\abovecaptionskip
  \sbox\@tempboxa{{\captionfonts #1: #2}}%
  \ifdim \wd\@tempboxa >\hsize
    {\captionfonts #1: #2\par}
  \else
    \hbox to\hsize{\hfil\box\@tempboxa\hfil}%
  \fi
  \vskip\belowcaptionskip}
\makeatother

\def\lsim{ \lower .75ex \hbox{$\sim$} \llap{\raise .27ex
\hbox{$<$}} }
\def\gsim{ \lower .75ex \hbox{$\sim$} \llap{\raise .27ex
\hbox{$>$}} }

\def\fnote#1#2{\begingroup\def\thefootnote{#1}\footnote{#2}
     \addtocounter{footnote}{-1}\endgroup}

\let\oldsqrt\sqrt
\def\sqrt{\mathpalette\DHLhksqrt}
\def\DHLhksqrt#1#2{%
\setbox0=\hbox{$#1\oldsqrt{#2\,}$}\dimen0=\ht0
\advance\dimen0-0.2\ht0
\setbox2=\hbox{\vrule height\ht0 depth -\dimen0}%
{\box0\lower0.4pt\box2}}

\begin{document}

\title{The Thin-Wall Approximation in Vacuum Decay: \\a Lemma}


\author{Adam~R.~Brown  \\
 \textit{\small{Physics Department, Stanford University, Stanford, CA 94305, USA}} }
\date{}
\maketitle
\fnote{}{\hspace{-.65cm}email: \tt{mr.adam.brown@gmail.com}}
\vspace{-.95cm}


\begin{abstract}
\noindent The `thin-wall approximation' gives a simple estimate of the decay rate of an unstable quantum field. Unfortunately, the approximation is uncontrolled.  In this paper I show that there are actually two different thin-wall approximations and that they bracket the true decay rate: I prove that one is an upper bound and the other a lower bound. In the thin-wall limit, the two approximations converge. In the presence of gravity, a generalization of this lemma provides a simple sufficient condition for non-perturbative vacuum instability. 
\end{abstract}

\thispagestyle{empty} 
\newpage 

Metastable states may decay by quantum tunneling. In the semiclassical (small $\hbar$) limit, the most important contribution to the decay rate
\begin{equation}
\textrm{rate} \sim A \exp[-B/\hbar],
\end{equation}
is the tunneling exponent $B$. In a famous paper Coleman explained how to calculate $B$ for a scalar field \cite{Coleman:1977py}. He showed that $B$ is given by the Euclidean action of an instanton that interpolates from the metastable (or `false') vacuum towards the target (or `true') vacuum. For a given field potential $V(\phi)$, the instanton can be derived by numerically integrating the second-order Euclidean equations of motion.

For those who lack either the computing power or the patience to solve the equations of motion, or who seek an intuitive understanding of the parametric dependence of the decay rate, Coleman also showed that we can often use the `thin-wall' approximation. While the instanton is a bubble that smoothly interpolates from the false vacuum to near the true vacuum, the thin-wall approximation treats this transition as abrupt \cite{Coleman:1977py}; the decay exponent is then approximated by
\begin{equation}
B \sim \bar{B}_\textrm{tw} \equiv \frac{27 \pi^2}{2} \frac{\sigma^4}{(V_\textrm{false} - V_\textrm{true})^3}. \label{eq:ambiguousthinwall}
\end{equation}
Consider these two expressions for the tension of the bubble wall $\sigma$,
\begin{equation}
\sigma_\textrm{min} \equiv \int^{\phi_\textrm{f}}_{\phi_*} d \phi \sqrt{2 (V[\phi] - V[\phi_\textrm{f}])} \ \ \ ; \ \ \ \sigma_\textrm{max} \equiv \int^{\phi_\textrm{f}}_{\phi_\textrm{t}} d \phi \sqrt{2 (V[\phi] - V[\phi_\textrm{t}])}. \label{eq:twodifferentsigmas}
\end{equation}
Coleman used only $\sigma_\textrm{min}$ and left Eq.~\ref{eq:ambiguousthinwall} as an uncontrolled approximation, in the sense that we are provided neither with an estimate of its accuracy nor with a bound on its error. We can do better. The main result of this paper is that, for any potential $V(\phi)$,
\begin{equation}
\hspace{-4cm} \textrm{Lemma 1 (no gravity)}: \ \  \boxed{ \ \bar{B}_\textrm{tw} [\sigma_\textrm{min}] \leq B \leq \bar{B}_\textrm{tw} [\sigma_\textrm{max}]  \ \T \B} \, . \label{eq:lemma1}
\end{equation}
The first inequality is proved in Appendix~\ref{subsec:B>Bmin}; the second in Appendix~\ref{subsec:Bmax>B}. As $V_\textrm{false} - V_\textrm{true} \rightarrow 0$ so too $\sigma_\textrm{max} - \sigma_\textrm{min} \rightarrow 0$ and the two thin-wall approximations converge.

\begin{figure}[htbp] 
   \centering
   \includegraphics[width=6in]{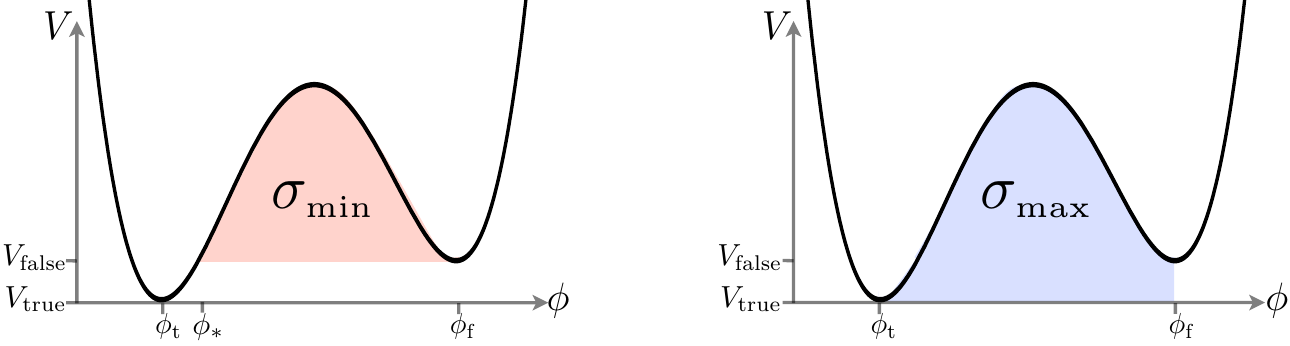} 
   \caption{The field starts in the false vacuum $\phi_\textrm{f}$ and tunnels towards the true vacuum $\phi_\textrm{t}$. The definition of $\phi_*$ is such that $V[\phi_*] = V[\phi_\textrm{f}]$. Equation~\ref{eq:twodifferentsigmas} gives two different definitions of the tension, with $\sigma_\textrm{max} > \sigma_\textrm{min}$.}
   \label{fig-sigmadefinitions}
\end{figure}

\newpage

 We can partially generalize this to include gravity. As first calculated by Coleman and de Luccia \cite{Coleman:1980aw},  gravitational backreaction changes $B$: the decay exponent now depends not just on the difference $\textrm{V}_\textrm{false} - \textrm{V}_\textrm{true}$, but also on $\textrm{V}_\textrm{false}$ and $\textrm{V}_\textrm{true}$ separately, since zero-point energy curves spacetime. The gravitational generalization of the thin-wall approximation to the decay exponent, $\bar{B}^G_\textrm{tw}[\sigma]$, is given by Eq.~\ref{eq:thinwallBwithgravity}, and 
the partial generalization of Lemma 1 is that
\begin{equation}
\hspace{-4cm} \textrm{Lemma 2 (including gravity, when $GV_\textrm{false} \leq 0$)}: \ \ \boxed{ \  B \leq \bar{B}^G_\textrm{tw} [\sigma_\textrm{max}] \ \B \T } \ . \label{eq:lemma2}
\end{equation}
This inequality is proved in Appendix~\ref{appendix:grav}. 
The addition of gravity means it is no longer always true that $B \geq\bar{B}^{G}_\textrm{tw}[\sigma_\textrm{min}]$: a de Sitter false vacuum ($G \, V_\textrm{false} > 0$) with a short but very broad barrier has arbitrarily large $\sigma_\textrm{min}$ but still decays relatively promptly via a Hawking-Moss instanton \cite{Hawking:1981fz}, so that $B \ll \bar{B}^G_{\textrm{tw}}[\sigma_\textrm{min}] < \bar{B}^G_{\textrm{tw}}[\sigma_\textrm{max}]$; and a Minkowski false vacuum ($G \, V_\textrm{false} = 0$) that has $B < \bar{B}^G_{\textrm{tw}}[\sigma_\textrm{min}] $ was exhibited numerically in Sec.~V of \cite{Samuel:1991dy}. (It is still open whether $B$ may be greater than $\bar{B}^G_\textrm{tw}[\sigma_\textrm{max}]$ for $G \, V_\textrm{false} > 0$;  I suspect it may not.)
\begin{figure}[htbp] 
   \centering
   \includegraphics[width=4in]{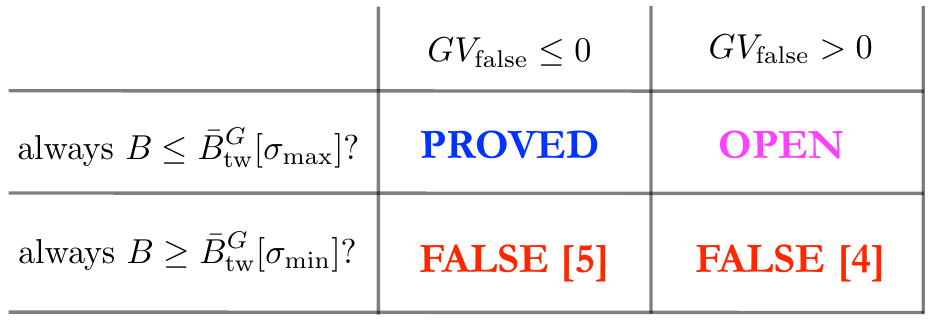} \hspace{1.3cm}
\end{figure}

One application of these lemmas is diagnosing instability. Gravity can stabilize 
superficially metastable $G \hspace{1pt} V_\textrm{false} \leq 0$ vacua \cite{Coleman:1980aw}, and there is great interest in determining which vacua decay and which endure (see e.g. \cite{Ooguri:2016pdq}). Lemma 2 provides a sufficient condition for instability, which is that $\bar{B}^G_{\textrm{tw}}[\sigma_\textrm{max}]  < \infty$, or equivalently
\begin{equation}
\textrm{sufficient condition for instability}: \ \ \sigma_\textrm{max} <  \frac{\sqrt{-V[\phi_\textrm{true}]} - \sqrt{-V[\phi_\textrm{false}]}}{\sqrt{6 \pi G}} .  \label{eq:sufficientcondition}
\end{equation} 
No such condition is required for  $G \hspace{1pt} V_\textrm{false} > 0$, since all de Sitter false vacua are unstable \cite{Coleman:1980aw}.

We can generalize Lemma 2 by replacing $\phi_\textrm{t}$ with any value in the range $\phi_\textrm{t} \leq \phi_\textrm{mid} < \phi_*$. Thus the false vacuum is unstable if there exists any  $\phi_\textrm{mid}$ such that
\begin{equation}
\sigma_\textrm{mid} \equiv   \int_{\phi_\textrm{mid}}^{\phi_\textrm{f}}  d \phi  \sqrt{2 (V[\phi] - V[\phi_\textrm{mid}])}  <  \frac{\sqrt{-V[\phi_\textrm{mid}]} - \sqrt{-V[\phi_\textrm{false}]}}{\sqrt{6 \pi G}} .  \label{eq:mostgeneral}
\end{equation}
When there are multiple fields, there are many possible routes over the barrier, and this condition applies to all of them: if we can find any route for the $\sigma_\textrm{mid}$ integral such that Eq.~\ref{eq:mostgeneral} holds, the vacuum must be unstable, the field must eventually decay, and spacetime is doomed \cite{Coleman:1980aw}.

\begin{figure}[htbp] 
   \centering
   \includegraphics[width=5.7in]{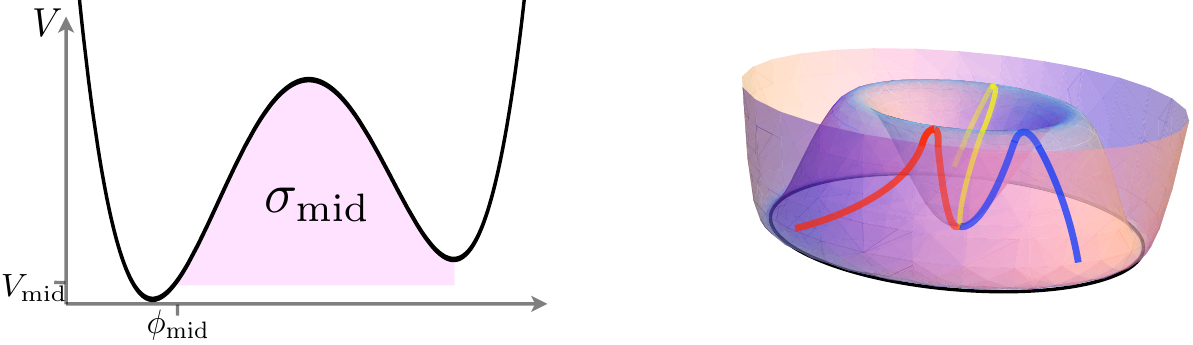}
\caption{If we can find \emph{any} path across the barrier that satisfies Eq.~\ref{eq:mostgeneral}, the vacuum must be unstable. Left: the $\sigma$ integral need not be taken all the way to $\phi_\textrm{true}$, and can instead stop anywhere in the range $\phi_\textrm{true} \leq \phi_\textrm{mid} < \phi_*$. Right: when there are multiple fields,
there are many escape routes, and the decay path may take any of them.}
   \label{fig-anystartandstop}
\end{figure}

\section*{Acknowledgements}
Thank you to Sonia Paban and Erick Weinberg for reading a draft of this paper, to Oliver Janssen for pointing out that I could drop the assumption that the gravitational instanton is O(4) symmetric, and to the organizers of TASI '09 for their hospitality while these lemmas were established.

\appendix

\section{Proving non-gravitational results} \label{appendix:nongrav}

This appendix proves Lemma 1. The two inequalities  demand two very different proof strategies. \\

\noindent Consider the decay of the false vacuum of Fig.~\ref{fig-sigmadefinitions}. We will assume that there are no intervening minima between the false and true vacua, 
but otherwise leave the potential completely general. I will now review how to calculate the vacuum decay rate; this is all explained with great clarity in \cite{Coleman:1977py}. In \cite{Coleman:1977py} it is shown that the tunneling exponent is given by the Euclidean action of an instanton. The instanton $\bar{\phi}(\tau,\vec{x})$ lives on $\mathbb{R}^4$ and extremizes the Euclidean action 
\begin{equation}
S_E = \int d^4x \left(   \frac{1}{2} (\partial_\mu \phi)^2 + V(\phi) \right) = 2 \pi^2  \int d\rho \, \rho^3 \left(   \frac{1}{2} \dot{\phi}^2 + V(\phi) \right) ,
\end{equation}
where we have written $ds^2 = d \tau^2 + d \vec{x}^2 = d \rho^2 + \rho^2 d \Omega_3^2$ and used that the instanton can be shown \cite{Coleman:1977kc} to have O(4) spherical symmetry $\bar{\phi} = \bar{\phi}(\rho)$. One boundary condition is that the field returns to the false vacuum $\bar{\phi} \rightarrow \phi_\textrm{f}$ as $\rho \rightarrow \infty$; the other boundary condition is that the field immediately after nucleation, given by 
\begin{equation}
\textrm{field after nucleation}: \ \  \phi(t=0,\vec{x}) = \bar{\phi}(\tau=0,\vec{x}) ,
\end{equation} 
has the same energy as the false vacuum ($\Delta E = 0$) and classically evolves towards the true vacuum. (In fact, the symmetry tells us that it will give rise to a bubble of approximately true vacuum that expands out at approaching the speed of light.) The instanton gives a path through the space of $\phi(\vec{x})$s that connects the before-tunneling configuration $\phi(\tau \hspace{-.5mm} = \hspace{-1mm} -\infty,\vec{x}) = \phi_\textrm{f}$ to the after-tunneling configuration $\phi(\tau=0,\vec{x})$. Indeed, the instanton is defined as the solution of minimum Euclidean action that satisfies these boundary conditions---in the language of \cite{Banks:1973ps}, it is the most probable decay path\footnote{Technically we insist that the decay path ends at its \emph{first} intersection with the $\Delta E=0$ surface.}. 

\vspace{1mm}

\noindent (With no constraint on the energy of the $\tau =0$ configuration, the instanton has exactly one negative mode \cite{Callan:1977pt}; this negative mode is associated with changing the energy of the nucleated bubble away from $\Delta E=0$. In this paper we will freeze this negative mode by only considering paths that end with the same energy as the false vacuum $\Delta E=0$; amongst this set of paths the instanton is a \emph{minimum} of the Euclidean action \cite{Coleman:1987rm}.)

\vspace{1mm}

\noindent To minimize the action, the instanton must satisfy the Euler-Lagrange equation,
\begin{equation}
\frac{d}{d\rho} \left( \frac{1}{2} \dot{\bar{\phi}}^2 - V(\bar{\phi}) \right) = -   \frac{3}{\rho} \dot{\bar{\phi}}^2 .\label{eq:frictioneom}
\end{equation}
This equation implies that $\frac{1}{2} \dot{\bar{\phi}}^2 - V(\bar{\phi})$ monotonically decreases as $\rho$ increases (in \cite{Coleman:1977py} this is called `friction'), which in turn implies that
\begin{eqnarray}
- V_\textrm{false} < & \frac{1}{2} \dot{\bar{\phi}}^2 - V(\phi) &< - V_\textrm{true} \\
\rightarrow \ \ \  \sqrt{2(V - V_\textrm{false})} <& \dot{\bar{\phi}}& < \sqrt{2(V - V_\textrm{true})} . \label{eq:boundonphidot}
\end{eqnarray}
The tunneling exponent is then given by the difference in Euclidean action between the instanton and the false vacuum
\begin{equation}
B = S_E(\bar{\phi}) - S_E(\bar{\phi}_\textrm{f})  =  2 \pi^2  \int_0^{\infty} d\rho \, \rho^3 \left(  \frac{1}{2} \dot{\phi}^2 + V(\phi) - V_\textrm{false}\right) . \label{eq:Bintermsofintegral}
\end{equation}
Since tunneling conserves energy,  immediately after nucleation the bubble must have the same energy as the false vacuum,
\begin{equation}
\Delta E \equiv 4 \pi \int_0^{\infty} d \rho \, \rho^2 \left( \frac{1}{2} \dot{\bar{\phi}}^2 + V(\bar{\phi}) - V_\textrm{false} \right)  = 0. \label{eq:DeltaE}
\end{equation}
But the energy density is not zero everywhere. Instead, the energy density is positive in the `wall' of the bubble where the field traverses the barrier, and then negative inside the bubble. 
\begin{figure}[htbp] 
   \centering
   \includegraphics[width=5in]{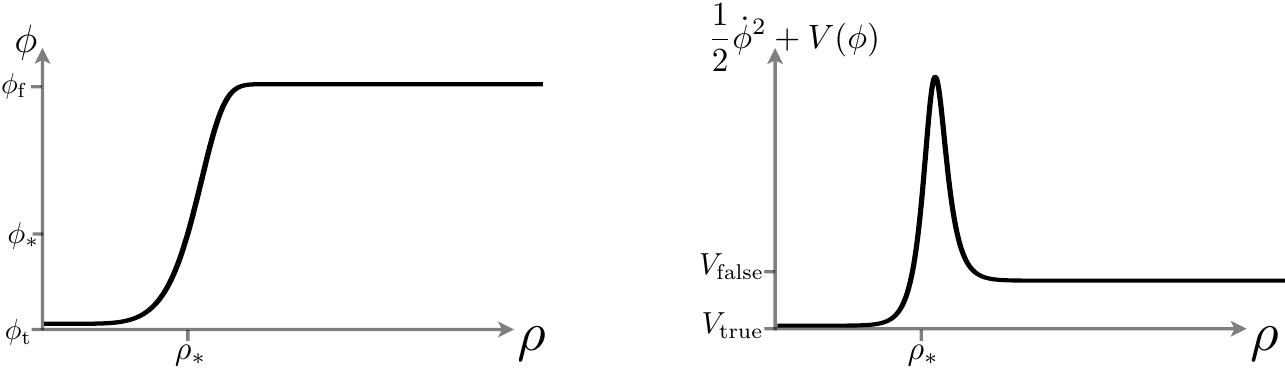}  
   \caption{A cross-section through a typical bubble at the moment of nucleation. The center of the bubble may have an energy density as low as $V_\textrm{true} $; but for $\rho > \rho_*$ the energy density necessarily exceeds that of the false vacuum $\frac{1}{2} \dot{\phi}^2 + V(\phi) \geq V(\phi) \geq V_\textrm{false}$. The total integrated energy is equal to that in the false vacuum, $\Delta E = 0$.}  \label{fig:crosssection}
\end{figure}
Fig.~\ref{fig:crosssection} shows a cross-section of a typical bubble. `Outside' the bubble ($\phi > \phi_*$) the energy density is bigger than $V_\textrm{false}$; `inside' the bubble ($\phi < \phi_*$) we know only that the energy density is bigger than $V_\textrm{true}$. \\

\noindent Consider the field value at the very center of the bubble, $\phi(\rho=0)$. It follows from the conservation of energy that $\phi(0) \leq \phi_{*}$; conversely it follows from Eq.~\ref{eq:boundonphidot} that $\phi(0) \geq \phi_\textrm{t}$. Indeed, the fact that $\phi_\textrm{t} \leq \phi(0) \leq \phi_*$ is what originally motivated the definitions of $\sigma_\textrm{min}$ and $\sigma_\textrm{max}$ in Eq.~\ref{eq:twodifferentsigmas}: the two lower limits of integration capture the full range of possible values of $\phi(0)$.\\

\noindent For the remainder of Appendix~\ref{appendix:nongrav} we will add a constant to the potential to set $V_\textrm{false} = 0$.

\subsection{Proving $B \geq \bar{B}_\textrm{tw}[\sigma_\textrm{min}]$} \label{subsec:B>Bmin}

Proof strategy: The instanton is a bubble configuration with zero energy, $\Delta E=0$. In order to have  $\Delta E = 0$, any bubble must be big, and if it is big enough it must have $B \geq \bar{B}_\textrm{tw}[\sigma_\textrm{min}]$. \\

\begin{figure}[htbp] 
   \centering
   \includegraphics[width=5in]{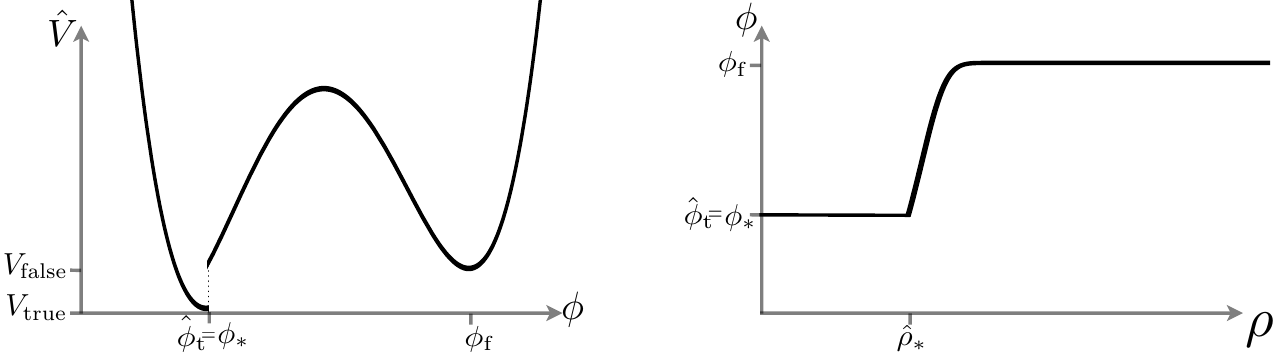} 
   \caption{$\hat{V}[\phi]$ is constructed by deleting the part of $V(\phi)$ that lies between $\phi_\textrm{t}$ and $\phi_*$. The corresponding bubble instanton has pure true vacuum inside some radius that we will call $\hat{\rho}_*$. The bubble has energy $\Delta \hat{E} = 0$.}
   \label{fig-Vhat}
\end{figure}

\noindent First let's construct a new potential $\hat{V}[\phi]$ that decays faster than $V[\phi]$,
\begin{displaymath}
\hat{V}[\phi] \equiv \left\{ \begin{array}{ll}
V[{\phi}] & \textrm{for $\phi>\phi_*$}\\
V[{\phi} - \phi_* + \phi_\textrm{t}] & \textrm{for $\phi<\phi_*$} \, \, .
  \end{array} \right.
\end{displaymath}
The region of $V[\phi]$ between $\phi_\textrm{t}$ and $\phi_*$ has been excised, with the $\phi > \phi_*$ part glued straight onto the $\phi<\phi_\textrm{t}$ part. This does not affect $\sigma_\textrm{min}$ or $V_\textrm{false}$ or $V_\textrm{true}$, but since we have removed part of the barrier the decay rate is faster
\begin{equation}
B[V(\phi)] \geq \hat{B} \equiv B[\hat{V}(\phi)].
\end{equation}
The shape of the bubble is plotted in Fig.~\ref{fig-Vhat}. The bubble stays uniformly in the true vacuum until $\rho=\hat{\rho}_*$, and then proceeds towards the false vacuum with a profile given by Eq.~\ref{eq:frictioneom}. 

\noindent Let's calculate $\hat{B}$. First notice that by changing variables from $\rho$ to $\phi$, 
\begin{equation}
\int_{\hat{\rho}_*}^{\infty} d\rho \left( \frac{1}{2} \dot{\bar{\phi}}^2  + V[\bar{\phi}]  \right) = \int_{\phi_*}^{\phi_\textrm{f}} \frac{d\phi}{\dot{\bar{\phi}}} \left( \frac{1}{2} \dot{\bar{\phi}}^2  + V[\bar{\phi}]  \right) = \int_{\phi_*}^{\phi_\textrm{f}} d \phi \frac{ \frac{1}{2} (\dot{\bar{\phi}} - \sqrt{2V})^2 +\sqrt{2V} \dot{\bar{\phi}}}{\dot{\bar{\phi}}} \geq \sigma_{\textrm{min}}. \label{eq:sigmacompletethesquare}
\end{equation}

\noindent The total energy of the bubble, relative to the false vacuum, is 
\begin{eqnarray}
\Delta \hat{E}   &=&  4 \pi \int_0^{\hat{\rho}_*} d\rho \rho^2 \left( \frac{1}{2} \dot{\bar{\phi}}^2 +  V[\bar{\phi}] 
\right) + 4 \pi \int_{\hat{\rho}_*}^{\infty} d\rho \rho^2 \left( \frac{1}{2} \dot{\bar{\phi}}^2  + V[\bar{\phi}] 
 \right)    \label{eq:Eintermsofintegral} \\
  & \geq & 4 \pi \int_0^{\hat{\rho}_*} d\rho \rho^2  V_\textrm{true} 
 + 4 \pi \hat{\rho}_*^2\int_{\hat{\rho}_*}^{\infty} d\rho  \left( \frac{1}{2} \dot{\bar{\phi}}^2  + V[\bar{\phi}] 
 \right) \nonumber \\
 & \geq  & \frac{4 \pi}{3} \hat{\rho}_*^3 \,V_\textrm{true} + 4 \pi \hat{\rho}_*^2 \, \sigma_\textrm{min}  \ .
 \end{eqnarray}
Tunneling conserves energy, $\Delta \hat{E} = 0$, so the bubble must be large
\begin{eqnarray}
 \hat{\rho}_*  & \geq  & \frac{3 \sigma_\textrm{min}}{(- V_\textrm{true})} .
 \end{eqnarray}
Since $\Delta \hat{E} = 0$ implies $\hat{B} = \hat{B} -  \frac{\pi \hat{\rho}_* \Delta \hat{E}}{2}$, we can use Eqs.~\ref{eq:Bintermsofintegral} \& \ref{eq:Eintermsofintegral} to prove our result
\begin{eqnarray}
B \ \geq \ \hat{B} & = & 2 \pi^2 \int_0^{\infty} d\rho \, \rho^2 ( \rho - \hat{\rho}_*) \left( \frac{1}{2} \dot{\bar{\phi}}^2 +  \hat{V}[\bar{\phi}] \right) \\
& \geq & 2 \pi^2 \int_0^{\hat{\rho}_*} d\rho \, \rho^2 ( \rho - \hat{\rho}_*) \left( \frac{1}{2} \dot{\bar{\phi}}^2 +  \hat{V}[\bar{\phi}] \right) \\
& \geq & 2 \pi^2 \int_0^{\hat{\rho}_*} d\rho \, \rho^2 ( \rho - \hat{\rho}_*) V_\textrm{true}  \\
& \geq &   2 \pi^2 \frac{\hat{\rho}_*^4 (-V_\textrm{true})}{12}  \\
& \geq & \frac{ 27 \pi^2}{2} \frac{ \sigma^4_\textrm{min}}{ (-V_\textrm{true})^3} \ \ = \ \  \bar{B}_\textrm{tw}[\sigma_\textrm{min}] \ .
\end{eqnarray}

\subsection{Proving $B \leq \bar{B}_\textrm{tw}[\sigma_\textrm{max}]$}  \label{subsec:Bmax>B}
Proof strategy: The instanton is the path of minimum action that interpolates from the false vacuum to a $\Delta E = 0$ state on the true vacuum side of the barrier. I will explicitly construct an interpolating path with action 
$\bar{B}_\textrm{tw}[\sigma_\textrm{max}]$. \\

\noindent Consider the one-parameter family of field-profiles\footnote{This field-profile has a discontinuous first derivative at both $\rho = \bar{R}$ and $\rho = 0$. This indicates the field-profile is not a minimum of the Euclidean action, but does not prevent the field-profile from contributing to the path integral---to contribute to the path integral, a field-profile only needs to be continuous, not differentiable. If desired,  my proof can be reformulated entirely in terms of differentiable field-profiles:  smoothing the field at $\rho = 0$ and $\rho = R$ will only make a tiny (second-order) change to the action, so the inequality would still follow.} $\phi_{\bar{R}} (\rho)$ parameterized by $\bar{R}$ and defined by
\begin{eqnarray}
\ \ \phi_{\bar{R}}(\rho) & = &\phi_\textrm{f} \ \ \ \  \ \ \ \ \ \ \ \  \ \  \ \ \ \ \   \hspace{1.63cm} \ \textrm{ for } \rho>\bar{R} \\
\dot{\phi}_{\bar{R}} (\rho) & =  &\sqrt{ 2(V[\phi_{\bar{R}}(\rho)] - V_\textrm{true})}  \  \ \ \ \ \   \textrm{ for } \rho<\bar{R} \ . \label{eq:definitionoffamily}
\end{eqnarray}
From the definition it follows that $\phi_\textrm{t} < \phi_{\bar{R}}(\rho)  \leq \phi_\textrm{f}$.  
 The Euclidean action of $\phi_{\bar{R}}(\rho)$ is 
\begin{eqnarray}
S_E[\bar{R}] & = & 2 \pi^2 \int_0^{\bar{R}} d\rho \, \rho^3 \biggl( \frac{1}{2} \dot{\phi}_{\bar{R}}^2 + V[\phi_{\bar{R}}]  \biggl) \\
& = & 2 \pi^2  \int_0^{\bar{R}} d\rho \, \rho^3 \biggl( 2(V[\phi] -  V_\textrm{true}) \biggl) + 2 \pi^2 \int_0^{\bar{R}} d\rho \, \rho^3  \biggl(V_\textrm{true} \biggl) \\
& \leq & 2 \pi^2  \bar{R}^3 \int_{\phi_{\bar{R}}[0]}^{\phi_\textrm{f}} d\phi \,   \sqrt{2(V[\phi] - V_\textrm{true})} + \frac{2 \pi^2 }{4} V_\textrm{true} \bar{R}^4 \\
& \leq & 2 \pi^2   \bar{R}^3 \sigma_{\textrm{max}}  - \frac{\pi^2 }{2} (-V_\textrm{true}) \bar{R}^4 \\
& \leq & \frac{27 \pi^2}{2} \frac{\sigma_{\textrm{max}} ^4}{(-V_\textrm{true}) ^3} \ \ = \ \ \bar{B}_\textrm{tw}[\sigma_\textrm{max}] \ .
\end{eqnarray}
Our family of field-profiles $\phi_{\bar{R}}(\rho)$ must contain as `escape path', in the language of \cite{Banks:1973ps}: while small values of $\bar{R}$ gives positive energy $\Delta E > 0$, arbitrarily large values of $\bar{R}$ give arbitrarily negative energies, so there must be an intervening value $\bar{R}_{E=0}$ such that $\Delta E[\phi_{\bar{R}}(t=0,\vec{x})] = 0$. Therefore the most probable escape path has $B \leq S_E[\bar{R}_{E=0}] \leq  \bar{B}_\textrm{tw}[\sigma_\textrm{max}]$. \\

\noindent This proof would still have gone through had we replaced $V_\textrm{true}$ by $V_\textrm{mid} \equiv V[\phi_\textrm{mid}]$ in Eq.~\ref{eq:definitionoffamily}, for any $\phi_\textrm{mid}$ in the range $\phi_\textrm{t} \leq \phi_\textrm{mid} < \phi_*$. Thus a strengthening of the lemma is that 
\begin{equation}
\ \ \ B \ \leq \ \frac{27 \pi^2}{2} \frac{\sigma_\textrm{mid}^4}{(V_\textrm{false} - V_\textrm{mid})^3}, \ \ \textrm{ where } \ \  \sigma_\textrm{mid} \equiv  \int_{\phi_\textrm{mid}}^{\phi_\textrm{f}}  d \phi  \sqrt{2 (V[\phi] - V[\phi_\textrm{mid}])} \label{eq:Bmid}
\end{equation}
for every possible $\phi_\textrm{mid}$, and (for multifield potentials) for every possible route over the barrier.

\section{Proving gravitational result} \label{appendix:grav}
When gravity is included, the decaying field curves spacetime. Despite this complication, the same general proof strategy will apply as for the non-gravitational case of Sec.~\ref{subsec:Bmax>B}. \\

\noindent The formalism that governs the gravitational decay of the false vacuum was described with great clarity in \cite{Coleman:1980aw}. The dominant \emph{known} instanton {is} $O(4)$-symmetric, and the metric may be written
\begin{equation}
ds^2 = d\xi^2 + \rho(\xi)^2  d \Omega_3^2. 
\end{equation}
Matter tells space how to curve
\begin{equation}
\dot{\rho}^2  = 1 + \frac{8 \pi G }{3} \rho^2 \left( \frac{1}{2} \dot{\phi}^2 - V(\phi) \right) , \label{eq:gravitationalconstraint}
\end{equation}
and space tells matter how to move
\begin{equation}
\frac{d}{d \xi} \left( \frac{1}{2} \dot{\bar{\phi}}^2 - V(\bar{\phi}) \right)  =  - 3 \frac{\dot{\rho}(\xi)}{\rho(\xi)} \dot{\bar{\phi}}^2 . \label{eq:gravityeom}
\end{equation}
When the gravitational constraint Eq.~\ref{eq:gravitationalconstraint} is satisfied, the action is given by Eq.~3.9 of \cite{Coleman:1980aw} as
\begin{equation}
S_E = 4 \pi^2 \int d \xi \left( \rho^3 V - \frac{3}{8 \pi G} \rho \right)  = 4 \pi^2 \int d \rho \frac{ \rho^3 V - \frac{3}{8 \pi G} \rho }{\dot{\rho}}  .
\end{equation}
The thin-wall approximation\footnote{For numerical investigations of the reliability of the thin-wall approximation in the non-gravitational case, see \cite{Samuel:1991mz}; see also \cite{Munster:1999hr}. For numerical investigations in the gravitational case, see \cite{Samuel:1991dy,Gen:1999gi}; see also \cite{Garfinkle:1989mv}.} to the tunneling exponent \cite{Coleman:1980aw,Parke:1982pm} is 
\begin{equation}
\bar{B}_\textrm{tw}^G \equiv 2 \pi^2 \bar{\rho}^3 \sigma +  \frac{3 }{16  } \frac{ (1 - \frac{8 \pi G}{3}  \bar{\rho}^2 V_\textrm{true} )^{\frac{3}{2}} - 1}{G^2 \, V_\textrm{true}} - \frac{3 }{16  } \frac{ (1 - \frac{8 \pi G}{3}  \bar{\rho}^2 V_\textrm{false} )^{\frac{3}{2}} - 1}{G^2 \, V_\textrm{false}},\label{eq:thinwallBwithgravity}
\end{equation}
where $\bar{\rho}$ is the radius of the bubble wall that maximizes Eq.~\ref{eq:thinwallBwithgravity}, namely (for $G V_f \leq 0$)
\begin{equation}
\bar{\rho} \equiv \frac{3 \sigma}{\sqrt{  \left( \sqrt{- V_\textrm{t}} - \sqrt{- V_\textrm{f}} \right)^2 - 6 \pi G \sigma^2 } \sqrt{ \left( \sqrt{- V_\textrm{t}} + \sqrt{- V_\textrm{f}} \right)^2 - 6 \pi G \sigma^2     }} .
\label{eq:thinwallrhowithgravity}
\end{equation}
As $\sigma$ approaches $(\sqrt{-V_\textrm{t}}  - \sqrt{-V_\textrm{f}} )/\sqrt{6 \pi G}$ both $\bar{\rho}$ and $\bar{B}_\textrm{tw}^G$ diverge; for larger values of the tension, the thin-wall approximation predicts that the false vacuum is stable. \\

\noindent To prove the gravitational lemma (Eq.~\ref{eq:lemma2}) we will follow Sec.~\ref{subsec:Bmax>B} in constructing a family of paths and showing that one member of the family is an escape path, and every member of the family has $\Delta S_E \leq \bar{B}^G_\textrm{tw} [ \sigma_\textrm{max}]$. The family of field-profiles will be parameterized by $\bar{R}$,
\begin{eqnarray}
 \phi_{\bar{R}}(\xi) & = &\phi_\textrm{f} \ \ \ \   \ \hspace{3.04cm}  \rightarrow  \  \dot{\rho}^2  =   1 - \frac{8 \pi G}{3}  \rho^2 V_\textrm{false}   \ \ \ \ \hspace{5mm}  \textrm{ for } \rho>\bar{R} \\
\dot{\phi}_{\bar{R}}(\xi) & =  &\sqrt{2(V[\phi(\xi)]  - V_\textrm{true})} \ \  \ \rightarrow \ \dot{\rho}^2  =  1 - \frac{8 \pi G}{3}  \rho^2  V_\textrm{true}   \ \ \ \ \ \  \ \  \textrm{ for } \rho<\bar{R} .
\end{eqnarray}
Notice that while this family need not satisfy the equation of motion for the field, Eq.~\ref{eq:gravityeom},  it is required to satisfy the gravitational constraint, Eq.~\ref{eq:gravitationalconstraint}, because only such configurations contribute to the gravitational path integral. Notice also that $\dot{\rho}\geq 1 > 0$ so $\rho$ monotonically increases with $\xi$ and the topology is $\mathbb{R}^4$. The field and metric differ from the false vacuum only inside $\rho < \bar{R}$, so the difference in action $\Delta S_E = S_E[\phi_{\bar{R}}(\rho)]  - S_E[\phi_\textrm{f}]$ is
\begin{eqnarray}
\Delta S_E  &=& 4 \pi^2 \int_0^{\bar{R}} d\rho \left(  \frac{ \rho^3 V - \frac{3}{8 \pi G} \rho }{{\sqrt{1 - \frac{8 \pi G}{3}  \rho^2 V_\textrm{true} }} } - \frac{ \rho^3 V_\textrm{false} - \frac{3}{8 \pi G} \rho }{{\sqrt{1 - \frac{8 \pi G}{3}  \rho^2 V_\textrm{false} }} }   \right) \\
& = & 4 \pi^2 \int_0^{\bar{R}} d\rho \left( \frac{ \rho^3 ( V - V_\textrm{true}) }{{\sqrt{1 - \frac{8 \pi G}{3}  \rho^2 V_\textrm{true} }} } +  \frac{ \rho^3 V_\textrm{true} - \frac{3}{8 \pi G} \rho }{{\sqrt{1 - \frac{8 \pi G}{3}  \rho^2 V_\textrm{true} }} } - \frac{ \rho^3 V_\textrm{false} - \frac{3}{8 \pi G} \rho }{{\sqrt{1 - \frac{8 \pi G}{3}  \rho^2 V_\textrm{false} }} }   \right)  \nonumber \\
& = & 4 \pi^2 \int^{\phi_\textrm{f}}_{\phi_{\bar{R}}[0]} \frac{d \phi \rho^3 (V-V_\textrm{t})}{\sqrt{2 (V - V_\textrm{t})}}  + \frac{3 }{16  } \frac{ (1 - \frac{8 \pi G}{3}  \bar{R}^2 V_\textrm{t} )^{\frac{3}{2}} - 1}{G^2 \, V_\textrm{t}} - \frac{3 }{16  } \frac{ (1 - \frac{8 \pi G}{3}  \bar{R}^2 V_\textrm{f} )^{\frac{3}{2}} - 1}{G^2 \, V_\textrm{f}} \nonumber \\
& \leq & 2 \pi^2 \bar{R}^3 \sigma_\textrm{max} +  \frac{3 }{16  } \frac{ (1 - \frac{8 \pi G}{3}  \bar{R}^2 V_\textrm{true} )^{\frac{3}{2}} - 1}{G^2 \, V_\textrm{true}} - \frac{3 }{16  } \frac{ (1 - \frac{8 \pi G}{3}  \bar{R}^2 V_\textrm{false} )^{\frac{3}{2}} - 1}{G^2 \, V_\textrm{false}} \label{eq:finalequation} \\
& \leq & \bar{B}_\textrm{tw}^G[\sigma_\textrm{max}] . 
\end{eqnarray}
First consider $\sigma_\textrm{max} < ({\sqrt{- V_\textrm{t}} - \sqrt{- V_\textrm{f}}})/{\sqrt{6 \pi G}}$. In this case, both $\Delta S_E [\bar{R}]$ and $\Delta E[\bar{R}]$ become unboundedly negative at large $\bar{R}$, so the family contains an escape path; since no member of the family has an action that exceeds $\bar{B}_\textrm{tw}^G[\sigma_\textrm{max}]$,  Lemma 2 holds. By contrast, for the case $\sigma_\textrm{max} \geq ({\sqrt{- V_\textrm{t}} - \sqrt{- V_\textrm{f}}})/{\sqrt{6 \pi G}}$ the family may not contain an escape path, but since $\bar{B}_\textrm{tw}^G[\sigma_\textrm{max}] = \infty$, Lemma 2 trivially holds. Thus we have proved Lemma 2 for all values of $\sigma_\textrm{max}$.\\

\begin{figure}[htbp] 
   \centering
   \includegraphics[width=6in]{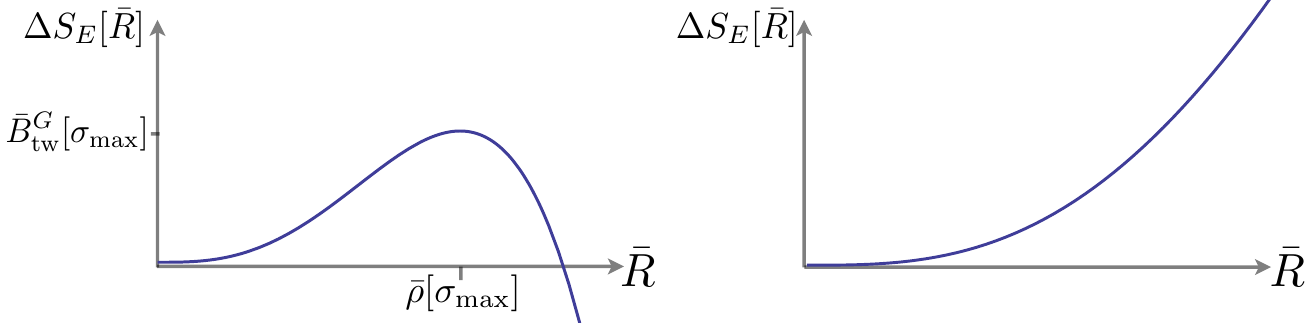} \ \ 
   \caption{Left: for $\sigma_\textrm{max} < ({\sqrt{- V_\textrm{t}} - \sqrt{- V_\textrm{f}}})/{\sqrt{6 \pi G}}$, the function $\Delta S_E [\bar{R}]$ from Eq.~\ref{eq:finalequation} never exceeds $\bar{B}^G_\textrm{tw}[\sigma_\textrm{max}]$.    \newline
    Right: for $\sigma_\textrm{max} \geq ({\sqrt{- V_\textrm{t}} - \sqrt{- V_\textrm{f}}})/{\sqrt{6 \pi G}}$, the function $\Delta S_E [\bar{R}]$ may (or may not) grow without bound.}
   \label{fig-twBofRHO}
\end{figure}

\noindent As in the non-gravitational case, this proof would still have gone through had we replaced $\phi_\textrm{t}$ with any value in the range $\phi_\textrm{t} \leq \phi_\textrm{mid} < \phi_*$, yielding the more powerful condition of Eq.~\ref{eq:mostgeneral}. \\

\noindent Nowhere in this proof did we need to assume that the dominant gravitational instanton has $O(4)$ symmetry. Instead, we used an $O(4)$-symmetric tunneling path to upper bound the action of the dominant instanton---the dominant instanton itself may have any symmetry or none.

\end{document}